\documentclass[twocolumn,showpacs,aps,pre,
groupedaddress,amssymb,amsmath,nobalancelastpage]{revtex4}

\usepackage{graphicx}
\usepackage{longtable}

\begin{document}

\title{Comparison of low amplitude oscillatory shear in experimental and
computational studies of model foams}

\author{Micah Lundberg$^1$, Kapilanjan Krishan$^{1,2}$, Ning Xu$^{3,4}$, Corey S. O'Hern$^{5,6}$ and Michael Dennin$^1$}
\affiliation{$^1$Department of Physics and Astronomy, University
of California at Irvine, Irvine, CA 92697-4575\\
$^2$ School of Physical Sciences Jawaharlal Nehru University, New Delhi 110067, India \\
$^{3}$Department of Physics and Astronomy, University of
Pennsylvania, Philadelphia, PA 19104-6396\\
$^{4}$James Franck Institute, The University of Chicago, Chicago,
IL 60637\\
$^5$Department of Mechanical Engineering, Yale University, New
Haven,
CT 06520-8286\\
$^6$Department of Physics, Yale University, New Haven, CT
06520-8120}
\date{\today}

\begin{abstract}
A fundamental difference between fluids and solids is their response
to applied shear.  Solids possess static shear moduli, while fluids do
not.  Complex fluids such as foams display an intermediate response to
shear with nontrivial frequency-dependent shear moduli.  In this
manuscript, we conduct coordinated experiments and numerical
simulations of model foams subjected to boundary-driven oscillatory,
planar shear. Our studies are performed on bubble rafts (experiments)
and the bubble model (simulations) in 2D.  We focus on the
low-amplitude flow regime in which T1 bubble rearrangement events do
not occur, yet the system transitions from solid- to liquid-like
behavior as the driving frequency is increased.  In both simulations
and experiments, we observe two distinct flow regimes. At low
frequencies $\omega$, the velocity profile of the bubbles increases
linearly with distance from the stationary wall, and there is a
nonzero total phase shift between the moving boundary and interior
bubbles. In this frequency regime, the total phase shift scales as a
power-law $\Delta \sim \omega^n$ with $n \approx 3$.  In
contrast, for frequencies above a crossover frequency $\omega >
\omega_{p}$, the total phase shift $\Delta$ scales linearly with the
driving frequency.  At even higher frequencies above a characteristic
frequency $\omega_{nl} > \omega_{p}$, the velocity profile changes
from linear to nonlinear. We fully characterize this transition from
solid- to liquid-like flow behavior in both the simulations and
experiments, and find qualitative and quantitative agreement for
the characteristic frequencies.
\end{abstract}

 \maketitle

\section{Introduction}

Aqueous foams are collections of gas bubbles that are separated by
liquid walls \cite{WH99}, and like other complex fluids, such as
pastes, emulsions, and granular media, they exhibit transitions
from solid- to liquid-like behavior in the response to applied
stress or strain.  For small strains, foams behave elastically
with stress proportional to strain.  Above the yield strain or
stress, bubble rearrangements occur and the system behaves as a
liquid.  In contrast to Newtonian fluids, foams display complex
spatio-temporal behavior in response to applied shear including
intermittency, shear banding, and nonlinear velocity profiles
\cite{CRBMGH02,VBBB03,KD03,LCD04,XOK05a,XOK05b,gsd06,jwh06,cjwh06,KMH08,CW08}.
Despite a number of experimental, numerical, and theoretical
studies of driven foams, a fundamental understanding of the
response of foam to applied shear is still lacking.

In this article, we describe a coordinated set of experimental and
numerical studies of model 2D foams undergoing applied oscillatory
planar shear to characterize the transition from solid- to
liquid-like behavior and from linear to nonlinear velocity
profiles.  There have been several studies of the response of foam
to steady shear, however, most of these have been performed in the
Couette geometry in which flow is confined between two concentric
cylinders \cite{LL87}.  Instead, we focus on planar shear to avoid
the `trivial' transition to nonlinear velocity profiles that stems
from the fact that in the Couette geometry the shear stress varies
with distance from the center of the shearing cell.

Another distinguishing feature of this work is its focus on {\it
oscillatory} rather than steady shear as the driving mechanism.
There are several reasons for selecting oscillatory shear.  First,
oscillatory shear allows one to control the amplitude
independently from the frequency of the driving.  When foams (and
other complex fluids) are driven by {\it steady} shear, they exist
in a highly fluidized state that is characterized by continuous,
often highly correlated bubble rearrangements, or T1 events
\cite{WH99}. In the highly fluidized state, the statistics of T1
events determine the flow curve and control stress fluctuations
\cite{HWB95,D97,JSSAG99,TSDKLL99,DK97,GD95,D04,OK95}. With
oscillatory shear, one can study the low-amplitude flow regime in
which particle rearrangement events do not occur, yet the system
can transition from solid- to liquid-like behavior and from linear
to nonlinear velocity profiles as the driving frequency is
increased. Since T1 events can be suppressed when using
oscillatory shear at low amplitude, the dissipation between fluid
films becomes the dominant relaxation mechanism \cite{dtgal08}.
Thus, in this regime one can directly probe the dissipation
mechanism by tuning the driving frequency.

In this article, we report on combined experiments and simulations
on model 2D foams: bubble rafts in experiments \cite{BL49} and the
bubble model in simulations \cite{D95}.  Bubble rafts consist of a
single layer of bubbles floating on the surface of water. (See
Fig.~\ref{bubble_raft} for a snapshot of the bubble raft used in
experiments.) Bubble rafts have a storied history as 2D model
systems for both crystalline and disordered solids
\cite{BL49,AK79}. In addition, we have performed a number of
studies characterizing these model systems by measuring and
quantifying T1 events \cite{LKXOD08}, stress fluctuations
\cite{PD03}, velocity profiles
\cite{LCD04,XOK05a,XOK05b,wkd06a,gsd06,wkd07}, and flow
transitions \cite{gsd06}. In this work, experiments on 2D bubble
rafts will be compared to simulations of the 2D bubble model
introduced by Durian~\cite{D95}. The bubble model treats foams as
soft disks that experience two pairwise forces when they overlap:
a repulsive linear spring force proportional to bubble overlap and
a dissipative force proportional to velocity differences between
bubbles.  A useful feature of the bubble model is that it can be
generalized to particles with finite mass \cite{XOK05a,XOK05b}.
Thus, the ratio of the damping and inertial forces can be varied
to interrogate the damping mechanism. Recent work has shown that
the bubble model successfully captures some of the features of the
dynamics of bubble rafts under shear, for example, the statistics
of individual T1 events \cite{LKXOD08}.  Thus, a comparison of
experiments of bubble rafts and simulations of the bubble model in
a well-controlled planar shear geometry will allow us to further
test under what conditions the bubble model accurately captures
the dynamics of model foams.

We will focus on measurements of the total phase shift between the
driving wall and interior bubble displacements, and velocity profiles
in systems subjected to low-amplitude oscillatory planar shear. At low
driving frequencies $\omega$, we observe a non-zero total phase shift,
while the velocity profiles rise linearly with distance from
the stationary wall. At low frequencies, the total phase shift scales as a
power-law $\Delta \sim \omega^n$ with $n \approx 3$.  In
contrast, for frequencies above a crossover frequency $\omega >
\omega_{p}$, the total phase shift $\Delta$ scales linearly with the
driving frequency. At even higher driving frequencies $\omega_{nl} >
\omega_{p}$, the velocity profiles transition from linear to
nonlinear. We compare the two crossover frequencies $\omega_{p}$ and
$\omega_{nl}$ in the experiments and simulations and find both
qualitative and quantitative agreement.  The structure of the
remainder of the manuscript will be organized as follows: section II,
theoretical background; section III, simulation methods; section IV,
experimental methods; section V, experimental and simulation results;
and section VI, conclusions.

\section{Theoretical Background}
\label{background}

\begin{figure}
\includegraphics[width=9cm]{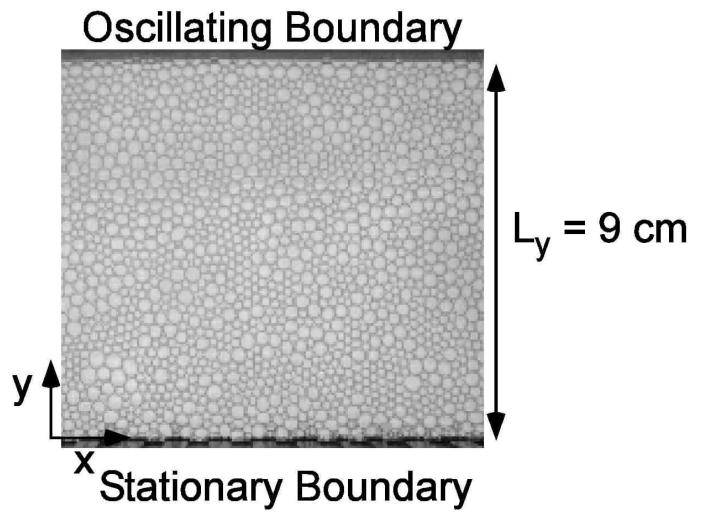}
\caption{A typical snapshot of the 2D bubble raft used in experiments with
packing fraction $\phi \approx 0.86$ and $N \approx 3700$ bubbles. The
bubble raft is composed of a bidisperse distribution of bubble sizes:
a $4$ to $1$ ratio of $2.5 \pm 0.3\ {\rm mm}$ to $5.3 \pm 0.5\ {\rm
mm}$ diameter bubbles.}
\label{bubble_raft}
\end{figure}

We will now review a simple theoretical treatment of the response of
an idealized viscoelastic material to an applied oscillatory strain,
which will provide a framework in which to interpret the experimental
and simulation results in Sec.~\ref{results}. The main point of this
section is to identify the possible flow regimes in viscoelastic
materials and their distinguishing properties. For illustration
purposes, we have selected the Kelvin-Voigt linear viscoelastic model
with frequency-independent elastic modulus and viscosity, though we
will discuss how the results from this model can be generalized.

If a planar oscillatory shear strain is applied to a viscoelastic
material (with shear along $x$ and shear gradient along $y$), the
$x$-displacement field $u_x(y)$ of the system relative to the
initial positions can be obtained by solving the force balance
equation \cite{landau}:
\begin{equation}
\frac{\partial \Sigma_{xy}}{\partial y}=\rho \frac{\partial
u^2_x(y)}{\partial t^2}, \label{cauchy}
\end{equation}
where the velocity field is $v_x = \partial u_x/\partial t$ and
$\rho$ is the areal mass density. The shear stress $\Sigma_{xy}$
includes both the elastic and viscous contributions.  As
mentioned, we will focus on a linear viscoelastic material with
\begin{equation}
\Sigma_{xy} = G\gamma + \eta \dot{\gamma},
\end{equation}
where the elastic contribution is proportional to the shear strain
$\gamma = \partial u_x/\partial y$ and the viscous contribution is
proportional to the shear rate $\dot{\gamma}= \partial v_x/\partial
y$.  $G$ is the elastic modulus, and $\eta$ is the dynamic viscosity
of the material. In general, complex fluids possess complex shear
moduli with arbitrary frequency dependence.  We have considered this
case, and find that the quantitative scaling of the crossover
frequency $\omega_p$ depends on the details of the viscoelastic model.
However, the qualitative features of the Kelvin-Voigt model, i.e. the
existence of $\omega_{p}$ and $\omega_{nl}$, are robust.

We consider the case of parallel plates aligned along the $x$-axis
separated by a distance $L_y$ in the $y$-direction as depicted in
Fig.~\ref{bubble_raft}.  The boundary at $y=0$ is stationary
$u_x(0,t)=0$ (bottom boundary), and the boundary at $y=L_y$ moves
according to $x_b(t) = u_x(L_y,t) = A \sin(\omega t)$ (top
boundary). The same geometry and notation is used for the
experiments and simulations.  To solve Eq.~\ref{cauchy}, we use
the ansatz $u_x(y,t) = A(y) \sin(\omega t)$ for the displacement
field.  Putting these elements together, we find the following
solution to Eq.~\ref{cauchy} for the displacement field
\begin{equation}
\label{dequation} u_x(y,t) = \mathrm{Im}\left[A e^{i\omega
t}\frac{\sin(ky)}{\sin(kL_y)}\right],
\end{equation}
and
\begin{equation}
v_x(y,t) = \mathrm{Im}\left[A i\omega e^{i\omega t}
\frac{\sin(ky)}{\sin(kL_y)}\right] \label{velequation}
\end{equation}
for the velocity field.  In Eqs.~\ref{dequation} and
\ref{velequation}, the wavenumber $k$ is complex, and satisfies
the dispersion relation
\begin{equation}
\omega^2 = \frac{G k^2}{\rho} + i \frac{\eta \omega k^2}{\rho}.
\end{equation}
or
\begin{equation}
\label{dispersion} k = (\omega \sqrt{\rho}) \frac{(G - i
\eta\omega)^{1/2}}{(G^2+(\eta \omega)^2)^{1/2}}
\end{equation}

Distinctive features of the velocity profile are best described by rewriting
Eq.~\ref{velequation} in terms of a $y$-dependent amplitude $v_{mag}(y)$
and local phase $\delta(y)$:
\begin{equation}\label{velampphase}
v_x(y,t) =  v_{mag}(y)\cos (\omega t - \delta(y)).
\end{equation}
We define the total phase shift $\Delta \equiv \delta(0) -
\delta(L_y)$. Because the flow is periodic, the velocity profile
at a given time $t$ repeats at subsequent times separated by
period $T = 2\pi/\omega$.  To simplify the analysis, we will focus
below on velocity profiles at times when the boundary velocity is
maximum ($t=0$ and $v_x(y,0)= v_{mag}(y)\cos(\delta(y))$). In the
simulations and experiments, statistical accuracy was improved by
averaging over driving cycles. We defined $v_x(y) \equiv \langle
v_x(y,2\pi p/\omega) \rangle_p$, where $\langle . \rangle_p$
indicates an average over $p$ cycles. Monitoring the full
time-dependence of the velocity profile is important, but is
outside the scope of the present work.  Error bars on the local
phase shift and velocity profile in the simulations and
experiments are given by the rms fluctuations within each bin and
are typically the size of the data points in the figures unless
otherwise noted.

\begin{figure}
\includegraphics[width=9.5cm]{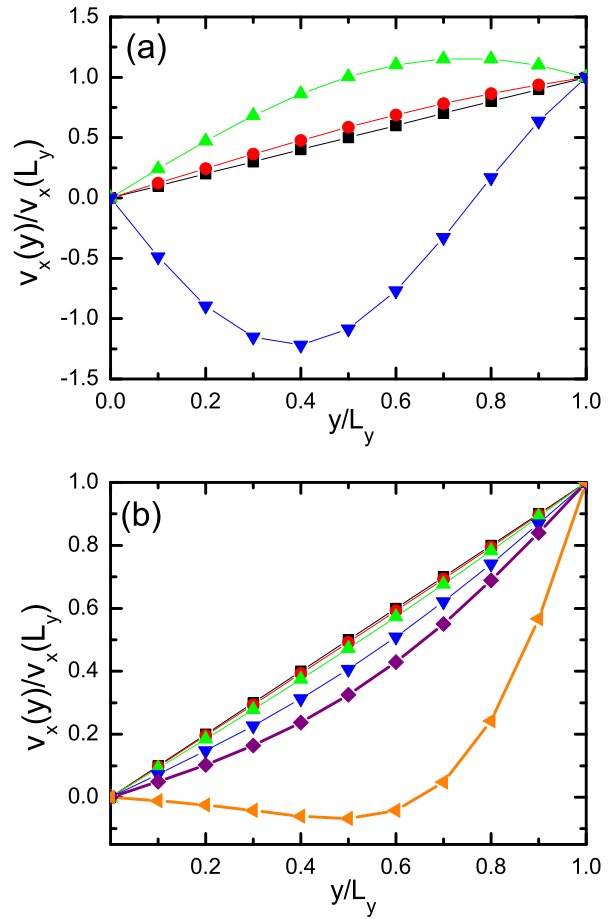}
\vspace{-0.4in} \caption{(Color online) Normalized horizontal
velocity profiles $v_x(y)/v_x(L_y)$ at $t = 0$ for the (a) pure
solid and (b) pure liquid obtained from solutions to
Eq.~\ref{cauchy} as a function of the driving frequency $\omega$.
We show $\omega/\omega_{nl} = 0.1$ (squares), $1.1.$ (circles),
$2.1$ (upward triangles), $4.1$ (downward triangles), $6.1$
(diamonds, only in (b)), and $20.1$ (left triangles, only in (b)).
When referring to the solid (liquid), $\omega_{nl}$ corresponds to
$\omega^s_{nl}$ ($\omega_{nl}^l$).} \label{theoryplots}
\end{figure}

It is instructive to consider two limits of the dispersion
relation in Eq.~\ref{dispersion}: the limit of a pure solid ($G
\ne 0$, $\eta = 0$) and the limit of a pure liquid ($G = 0$, $\eta
\ne 0$). For the pure solid, we recover the dispersion relation
$\omega/k = \sqrt{G/\rho} \equiv v_s$, where $v_s$ is the speed of
shear waves in the solid.  In this case, $k = \omega/v_s$ is real,
and the velocity field is a standing wave given by
\begin{equation}
\label{standing_wave} v_x(y,t) = A \omega \cos(\omega
t)\frac{\sin(\omega y/v_s)}{\sin(\omega L_y/v_s)}.
\end{equation}
For $\omega \ll \omega^s_{nl} \equiv v_s/L_y$, $\sin(\omega y/v_s)
\approx \omega y/v_s$, and the velocity profile becomes linear in
$y/L_y$, $v_x(y,t) \approx A \omega \cos(\omega t) y/L_y$.  For
fixed system size, the transition from linear to nonlinear
velocity profiles occurs when $\omega > \omega^s_{nl}$.  Because
$k$ is real for the case of the pure solid, the total phase shift
$\Delta = 0$, and the system oscillates in phase with the driving
wall.

In the limit of the pure liquid, we recover the dispersion relation $i
\omega = -\nu k^2$, where $\nu = \eta /\rho$ is the kinematic
viscosity. In this case, $k = (1-i)/D$ with $D = \sqrt{2\eta/(\omega
\rho)}$. The form of the velocity profile is more complex than that
for the pure solid. However, for small driving frequencies the
velocity profile can be expanded in powers of $L_y/D$, and the first
term is linear in $y/L_y$. Thus, for $L_y/D < 1$ or $\omega < 2
\omega^l_{nl}$, where $\omega^l_{nl} = \eta/(\rho L_y^2)$, the
velocity profiles are approximately linear, as we found for the pure
solid.  However, in contrast to the pure solid, there is a non-zero
total phase shift $\Delta$ in the pure liquid since the wavenumber $k$
is complex. The full form of the phase shift is complicated, but at
low driving frequencies one can expand $\Delta$ about $\omega=0$.  For
the pure liquid, at lowest order in $\omega$, we find that the total
phase shift scales linearly with the frequency, $\Delta \sim \omega/6
\omega^l_{nl}$.

\begin{figure}
\includegraphics[width=9cm]{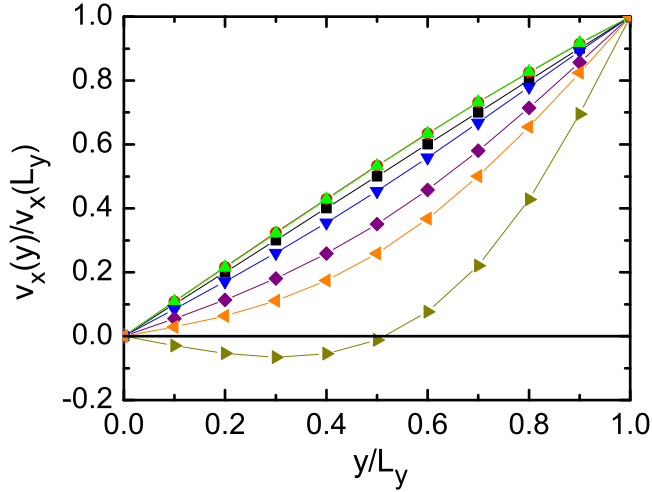}
\caption{(Color online) Normalized horizontal velocity profiles
$v_x(y)/v_x(L_y)$ at $t = 0$ for a viscoelastic material with $\beta = G \rho
L_y^2/\eta^2= 1$ obtained from solutions to Eq.~\ref{cauchy} as a
function of the driving frequency $\omega$. We show
$\omega/\omega_{nl} = 0.1$ (squares), $1.1.$ (circles), $2.1$
(upward triangles), $4.1$ (downward triangles), $6.1$ (diamonds),
$8$ (left triangles) and $20$ (right triangles). $\omega_{nl}$
corresponds to $\omega_{nl}^l = \omega_{nl}^s$. }
\label{theorymixedplot}
\end{figure}

In Fig.~\ref{theoryplots}, we plot the velocity profiles that
satisfy Eq.~\ref{cauchy} for two cases: (a) the pure solid ($G>0$
and $\eta = 0$) and (b) the pure liquid ($G=0$ and $\eta > 0$). In
Fig.~\ref{theoryplots}(a), we show that the velocity profiles for
the pure solid become nonlinear when $\omega/\omega^s_{nl} > 1$.
Note that the profiles first become nonlinear by developing
negative curvature above the linear profile, and then as the
frequency is increased further the profile develops positive
curvature below the linear profile. This nonmonotonic behavior is
caused by the standing wave solution in Eq.~\ref{standing_wave}.
In Fig.~\ref{theoryplots}(b), the velocity profile for the pure
liquid begins to deviate from a linear profile when $\omega > 2
\omega^l_{nl}$.  In contrast to the pure solid, the liquid system
only exhibits monotonically decaying nonlinear velocity profiles
with a `decay length' that decreases continuously with increasing
driving frequency. An interesting feature of these profiles is
that one can observe negative velocities at sufficiently high
frequencies as in the case of the pure solid.

We now consider the solution to Eq.~\ref{cauchy} for the velocity
profile in the more general case of a viscoelastic material with
nonzero $G$ and $\eta$.  In Fig.~\ref{theorymixedplot}, we show an
expanded range of driving frequencies for a viscoelastic material
with $\beta \equiv (\omega^s_{nl}/\omega^l_{nl})^2 = G \rho
L_y^2/\eta^2 = 1$, so that $\omega_{nl}^s = \omega_{nl}^l$.
Similar to the case of the pure solid, when $\omega \gtrsim
\omega^s_{nl}$ the velocity profiles show a small negative
curvature with the profile slightly above the linear profile at
$\omega/\omega^s_{nl} = 0.1$. At higher frequencies, the system
behaves similar to the pure liquid with monotonically decaying
profiles and a continuous decrease in the decay length with
increasing frequency.  At sufficiently high frequencies, we also
observe a regime in which the velocity becomes negative. For
viscoelastic materials, we define $\omega_d
> \omega_{nl}$ as the frequency above which the system begins to
display monotonically decaying `liquid-like' velocity profiles.

\begin{figure}
\includegraphics[width=9cm]{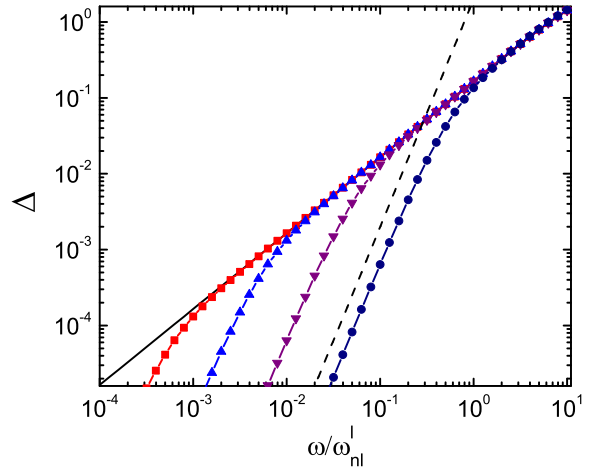}
\caption{(Color online) The total phase shift $\Delta = \delta(0)-
\delta(L_y)$ for viscoelastic materials with different values of
$\beta = G \rho L_y^2/\eta^2$ versus the normalized driving frequency
$\omega/\omega^l_{nl}$. The curves for all $\beta$ scale linearly with
$\omega$ at sufficiently high frequencies. We include $\beta=0$ (solid
line), $0.0005$ (squares), $0.005$ (upward triangles), $0.05$
(downward triangles), and $0.5$ (filled circles). The crossover
frequency $\omega_p$ can be obtained by locating the intersection of
the low-frequency and high-frequency power-law scaling forms for
$\Delta$. The slope of the black solid (dashed) line is $1$ ($3$).}
\label{theoryphaseplot}
\end{figure}

In Fig.~\ref{theoryphaseplot}, we show the total phase shift
$\Delta$ for the pure liquid and viscoelastic materials as a
function of the driving frequency. (The pure solid is not shown
since $\Delta$ is identically zero for all frequencies.) As
expected, $\Delta$ for the pure liquid ($\beta=0$) scales linearly
with $\omega$. For viscoelastic materials with $\beta > 0$, there
is a clear crossover from low-frequency scaling $\Delta \sim
\omega^n$ to high-frequency scaling $\Delta \sim \omega^m$ with
$m<n$.  For the Kelvin-Voigt model, the high frequency limit is
equivalent to $G = 0$, and $\Delta = \omega/6 \omega^l_{nl}$,
corresponding to $m=1$, and the low frequency limit is $\Delta =
\omega^3 \omega^l_{nl}/(6 {\omega^s_{nl}}^4)$, corresponding to $n
= 3$. Using these expressions, one can derive the crossover
frequency explicitly, $\omega_p = {\omega^s_{nl}}^2/\omega^l_{nl}
= \omega^l_{nl} \beta$. For a more general model with a complex
shear modulus, where the stress is given by $\Sigma_{xy} =
G^*(\omega)\gamma$, $G^*(\omega) = G'(\omega) + iG''(\omega)$, and
$G'$ and $G''$ are the storage and loss moduli, the values of $n$
and $m$ depend on the frequency dependence of $G'$ and $G''$.
However, for physically motivated $G^*(\omega)$, the crossover
from the low-frequency elastically dominated to high-frequency
viscously dominated behavior persists. One consequence of the
crossover in frequency dependence is that the low-frequency total
phase shift tends to zero rapidly at low frequencies, and thus
$\Delta$ may be difficult to measure at low frequencies in
experiments. In our experimental studies, we were able to detect
the change in scaling behavior of $\Delta$, but were not able to
measure the scaling exponents accurately. Much more sensitive
experiments are planned to measure $\omega_p$ and the storage and
loss moduli at low frequencies.

\begin{figure}
\includegraphics[width=9cm]{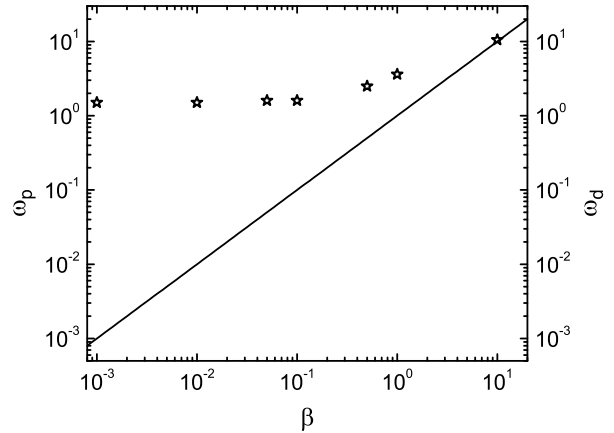}
\caption{Plot of $\omega_p$ (solid line) and $\omega_d$ (stars)
versus $\beta = G \rho L_y^2/\eta^2$ for the viscoelastic
Kelvin-Voigt model.  Note that $\omega_{p} < \omega_d$ over a wide
range of $\beta$.} \label{crossover}
\end{figure}

The dependence of $\omega_p$ and $\omega_d$ on $\beta$ for the
viscoelastic Kelvin-Voigt model is given in Fig.~\ref{crossover}.
Here we used the analytical result for $\omega_p$, but $\omega_d$
is determined numerically. This figure illustrates an important
feature of the model: for $\beta < 1$, $\omega_d$ is relatively
insensitive to $\beta$, i.e. whether the system is solid or
liquid, while $\omega_p$ decreases linearly with $\beta$. Thus, as
$\beta \rightarrow 0$, $\omega_p/\omega_d \rightarrow 0$. This is
consistent with the fact that as the system becomes more
solid-like, the initial deviations from nonlinearity are positive,
so the transition to liquid-like behavior is delayed to higher
frequencies. We expect similar behavior for more general models.
By measuring these characteristic frequencies, future experimental
studies will be able to characterize the material properties of
foams and other complex fluids.

It is helpful to summarize the three characteristic
frequencies---$\omega_p$, $\omega_{nl}$, and $\omega_d$---that
were defined above. $\omega_{p}$ is the crossover frequency that
characterizes the change in the scaling behavior of the total
phase shift as a function of frequency. For pure solids,
$\omega_p$ is not defined, for pure liquids, $\omega_p = 0$, and
for viscoelastic materials $\omega_p > 0$. $\omega_{nl}$ is the
frequency above which we observe deviations from linear behavior
in the horizontal velocity profile. For $\omega > \omega_{nl}$,
pure liquids display decaying nonlinear velocity profiles with
positive curvature, and decay more strongly with increasing
frequency. In solids and viscoelastic fluids, when $\omega >
\omega_{nl}$, the horizontal velocity profile initially possesses
negative curvature with deviations `above' the linear velocity
profile.  Thus, the curvature of the profile at low frequencies
near $\omega_{nl}$ can be used to differentiate `liquid-like' from
`solid-like' velocity profiles. In the experiments, the solid-like
response of the system at low frequencies is weak. Thus, we focus
on measuring $\omega_d$, the frequency above which the system
begins to display decaying `liquid-like' velocity profiles
(instead of $\omega_{nl}$) and $\omega_p$ in the simulations and
experiments.

\section{Simulation Methods}

We performed numerical simulations in 2D of the bubble model, which
was generalized to include particles with nonzero mass $m$, undergoing
boundary-driven, oscillatory planar shear flow.  The systems were
composed of a total of $N_t=1280$ disks with the same mass. Half of
the disks were small and the other half were large with diameter ratio
$r=2$ to avoid crystallization under shear and match the bubble
distribution used in the bubble raft experiments.  The original
simulation cell was rectangular with $x$-coordinates in the range
$[0,L_x]$ and $y$-coordinates in the range $[-L_y/8,9 L_y/8]$.
Bubbles were initialized with random initial positions within this
rectangular domain at a given packing fraction $\phi$ and then the
system was relaxed to the nearest local potential energy minimum using
conjugate gradient energy minimization \cite{num} with periodic
boundary conditions in both the $x-$ and $y-$directions.  All bubbles
outside $y=[0,L_y]$ formed two rough, rigid boundaries. Bubbles with
$y$-coordinates in the range $[L_y,9 L_y/8]$ ($[-L_y/8,0]$) formed the
top (bottom) boundary.  $N \approx 1000$ disks filled the interior of
the cell between the two rigid boundaries.  After the top and bottom
boundaries were formed, we used periodic boundary conditions only in
the $x$-direction.  Packing fractions in the range $\phi=[0.85,0.9]$
were investigated.

In the bubble model, bubble $i$ experiences two pairwise forces from
neighboring bubbles $j$ that overlap $i$: 1. repulsive linear spring
forces that arise from bubble deformation
\begin{equation}
\label{repulsive} {\vec F}^r_{ij} = \frac{\epsilon}{\sigma_{ij}}
\left(1-\frac{r_{ij}}{\sigma_{ij}}\right) {\hat r}_{ij},
\end{equation}
and 2. viscous damping forces proportional to
the relative velocity between bubbles that arise from dissipation
between the fluid walls
\begin{equation}
\label{viscous} {{\vec F}^v}_{ij} = -b \left( {\vec v}_i - {\vec
v}_j \right),
\end{equation}
where $\epsilon$ sets the energy scale for elastic deformation,
$\sigma_{ij} = (\sigma_i + \sigma_j)/2$ is the average diameter,
$r_{ij}$ is the center-to-center separation between bubbles $i$ and
$j$, ${\hat r}_{ij} = {\vec r}_{ij}/r_{ij}$ is the unit vector that
points from the center of bubble $j$ to the center of bubble $i$, and
$b$ is the damping coefficient.  Note that when bubbles $i$ and $j$ do
not overlap, the pairwise forces ${\vec F}^r_{ij}={{\vec
F}^v}_{ij}=0$.

The ratio of the damping to inertial forces can be expressed via a
dimensionless damping coefficient $b^* = b \sigma/\sqrt{\epsilon
m}$, where $\sigma$ is the small bubble diameter.  Underdamped
(overdamped) systems are characterized by $b^* < b^*_c$ ($b^*
>b^*_c$), where $b^*_c = \sqrt{2}$ for linear spring interactions.
We studied both under and overdamped systems in the range
$b^*=[0.1,3]$.  The units of energy, length, and time in the
simulations are $\epsilon$, $\sigma$, and
$\sigma/\sqrt{\epsilon/m}$, respectively.

The time evolution of the position ${\vec r}_i$ and velocity
${\vec v}_i$ of an interior bubble $i$ can be
obtained by integrating the equation of motion
\begin{equation}
\label{eom}
m\frac{d^2 \vec{r}_i}{dt^2} = \sum_j \left( \vec{F}^r_{ij} + \vec{F}^v_{ij}
 \right).
\end{equation}
We employed standard Gear predictor-corrector algorithms to
numerically integrate Eq.~\ref{eom} for the positions and velocities of
the interior bubbles~\cite{allen}.  This simple `discrete element',
short-range model for 2D foams allows us to quickly and efficiently
generate an ensemble of configurations with a given set of external
boundary conditions.

Oscillatory planar shear was imposed by rigidly moving all of the
bubbles that comprise the top boundary in the $x$-direction as a
function of time according to
\begin{equation}
\label{boundary}
x_b(t) = A \sin(\omega t),
\end{equation}
while bubbles that comprise the bottom boundary remain stationary.
$A=0.8\sigma$ and $\omega$ are the amplitude and angular frequency of
the sinusoidal driving.  At this amplitude, we did not observe any T1 events
\cite{LKXOD08} over the entire range of driving frequencies studied.

\begin{figure}
\includegraphics[width=8.8cm]{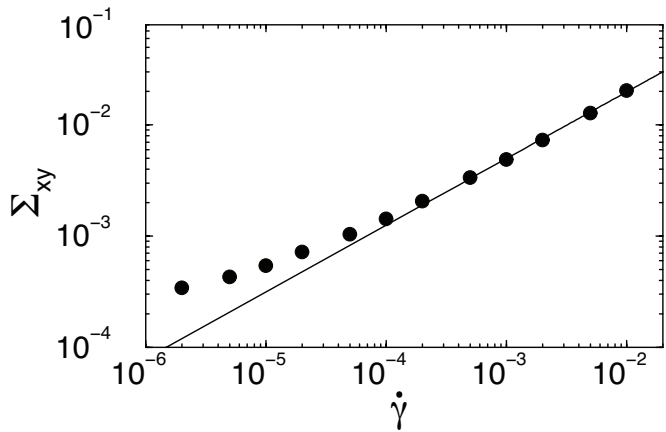}
\caption{Average shear stress $\Sigma_{xy}$ plotted versus the applied
shear rate ${\dot \gamma}$ for a system with $\phi=0.86$ and $b^*=2$
undergoing steady planar shear flow in 2D simulations of the bubble model.
The crossover shear rate at which the system transitions
from quasistatic to power-law behavior is ${\dot \gamma}_c \approx
10^{-4}$.}
\label{flowcurve}
\end{figure}

We calculated several important physical quantities in the
simulations, including the phase shift of bubble $x$-displacements
relative to the motion of the boundary and the horizontal velocity
profiles of the bubbles.  The $x$-displacements and velocity profiles
reached steady state after a few cycles; thus, we began measurements
after $5$ cycles and continued for an additional $15$ cycles to
calculate averages.  The height dependence of the phase shift and
velocity profiles were measured by partitioning the simulation cell
into equal-sized rectangular bins centered at $y$ with height $\Delta
y \approx 2$ large particle diameters, with $y$ measured from the
bottom stationary boundary.

To calculate the local phase shift $\delta(y)$, we averaged the bubble
$x$-displacement $u_x(y,t)$ relative to the initial position over all
bubbles within the bin located at $y$.  We then fit the average bubble
$x$-displacement in each bin to $u_x(y,t) \sim \sin(\omega t -
\delta(y))$ to determine the local phase shift $\delta(y)$.  We
measured $\delta(y)$ at several times during a given cycle to verify
that it was time-independent, and the bubble motion was periodic.  To
measure the average horizontal velocity profile $v_x(y,t)$ of the
interior bubbles, we used a binning procedure identical to that
employed to measure $\delta(y)$.

An important characteristic time (or frequency) scale in the bubble
model is the shear rate ${\dot \gamma}_c$ at which the system
transitions from quasistatic behavior at low shear rate (shear stress
$\Sigma_{xy} \propto {\dot \gamma}^0$) to highly fluidized behavior at
high shear rate ($\Sigma_{xy} \propto {\dot \gamma}^\alpha$, where
$\alpha > 0$) when the system is driven by steady planar shear
\cite{D97}.  This frequency scale has also been measured in the bubble
raft experiments, and thus $\omega_c = 2 \pi {\dot \gamma}_c$ can be
used to normalize the crossover frequencies obtained in experiments
and simulations. To simulate systems undergoing steady planar shear,
we employed the same boundary-driven method as described above except
$x_b(t) = L_y {\dot \gamma} t$ instead of Eq.~\ref{boundary}. The flow
curve for a system with $\phi=0.86$ and $b^*=2$ is shown in
Fig.~\ref{flowcurve}, where the virial expression including
dissipative forces was used to calculate the shear stress
\cite{allen}. To determine ${\dot \gamma}_c$, we calculated the median
of the data point at which the flow curve first deviates by more than
$10\%$ from the power-law behavior and the previous data point at
higher shear rate.  For the flow curve in Fig.~\ref{flowcurve}, we
estimate ${\dot \gamma}_c \approx 2 \times 10^{-4}$.  ${\dot
\gamma}_c$ was determined for each value of $b$ and $\phi$.

\section{Experimental Methods}
\label{expmeths}

\begin{figure}
\includegraphics[width=9cm]{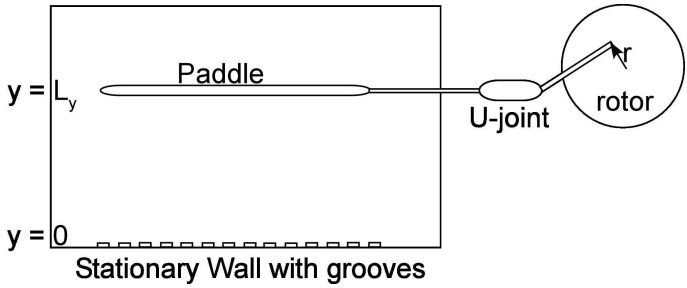}
\caption{A schematic of the experimental setup that applied
oscillatory planar shear to the bubble rafts.  The U-joint highlighted
in the figure is used to convert the rotary motion of the driving
motor into oscillatory, planar shear.}
\label{schematic}
\end{figure}

The experimental setup to apply oscillatory planar shear to bubble
rafts includes three components: a rectangular basin, oscillating
paddle, and imaging system.  A schematic of the experimental setup
is shown in Fig.~\ref{schematic}.  The basin has dimensions $37$
cm by $15$ cm and was filled to a depth of $5$ cm with a
surfactant solution. A paddle was located in the middle of the
basin, leaving a span of $9$ cm between it and the opposite wall.
As illustrated in the schematic in Fig.~\ref{schematic}, the ends
of the system are ``open'' in the following sense. The entire
basin is filled with bubbles, and the paddle only spans the
central portion of the system.  Furthermore, only the central
third of the bubbles in the region covered by the paddle are used
in the data analysis, and thus edge effects are minimized. The
paddle was driven by an M-drive $23$ stepper motor located outside
the basin. A rotor and universal (U-) joint were used to convert
the axial drive of the motor into linear sinusoidal motion.  The
amplitude of oscillation was varied by changing the contact point
between the rotor and U-joint.

The bubbles were confined between a movable paddle and a fixed
wall. The bubbles were constrained to move with the paddle using metal
tabs that extended one bubble diameter into the system. The tabs were
spaced approximately every 5 bubbles. The fixed wall consisted of a
series of square indentations approximately the size of a bubble. This
fixed the bubble velocities at the stationary wall to zero. Because
the first row of bubbles at each wall is interspersed with elements to
hold it in place, slight distortions of the bubbles prevented accurate
measurement of their positions. Therefore, in the experiments we
defined the location of $y = 0$ and $y = L_y$ to be the boundary
between the first and second rows of bubbles at the stationary wall
and paddle, respectively, instead of the location of the physical
boundaries. For a sinusoidally oscillating paddle that is initially
undisplaced, this gives the boundary condition at $y=L_y$: $x_b(t) = A
\sin(\omega t)$, where $A=0.8\sigma$ and $\omega$ are the amplitude
and frequency of the driving. At this amplitude, no T1 events were
recorded over the entire range of $\omega$.

The bubble raft consisted of a bidisperse distribution of bubble
sizes: a $4$ to $1$ ratio of $2.5 \pm 0.3\ {\rm mm}$ to $5.3 \pm
0.5\ {\rm mm}$ diameter bubbles, which corresponds to a diameter
ratio $r\approx 2.1$.  The solution composition was 80\%\
deionized water, 15\%\ glycerol, and 5\%\ miracle bubble (Imperial
Toy Corp), which is a commercially available surfactant.  The
bubbles were produced by passing compressed nitrogen through the
solution via a needle. The diameter of the needle and nitrogen
pressure determine the final size of the bubbles.  To create
bidisperse bubble mixtures, we used two needles with different
sizes at constant pressure. After the bubbles were produced, we
stirred the solution with a glass rod to break-up large-scale
crystalline domains so that only short-range order persisted.  We
tuned the pressure to prevent multiple layers of bubbles from
building up in the $z$-direction.

\begin{figure}
\includegraphics[width=9cm]{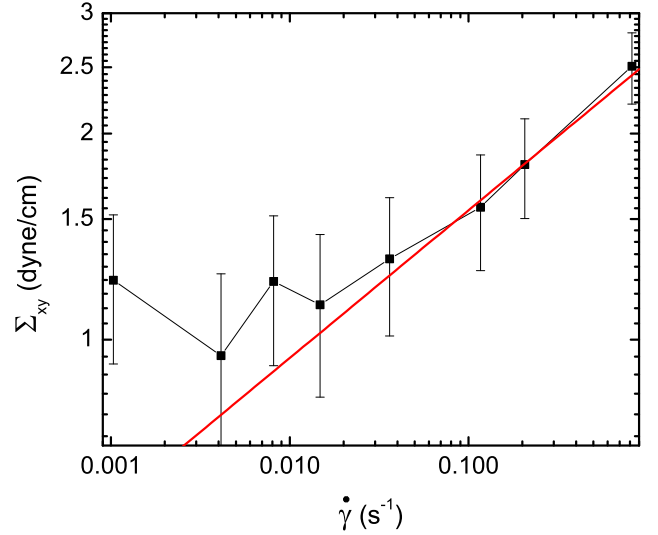}
\caption{(Color online) Shear stress $\Sigma_{xy}$ versus the
applied shear rate ${\dot \gamma}$ for the same bubble raft system
described in Fig.~\ref{bubble_raft} undergoing steady planar
shear.  We estimate the crossover frequency ${\dot \gamma}_c
\approx 0.065 {\rm s}^{-1}$ above which the system transitions
from the quasistatic to the power-law flow regime.}
\label{expflowcurve}
\end{figure}

The definition of the gas (or liquid) area fraction for a bubble raft
is somewhat imprecise because the bubbles form 3D structures on the
water surface, which makes it difficult to define the amount of fluid
in the walls. Also, using definitions based on T1 events present
challenges because true vertices do not exist in bubble rafts, and
therefore, minimum edge lengths are not well-defined. This is in
contrast to fully confined two-dimensional systems, such as soap
films, for which more precise definitions of liquid area fraction
exist \cite{RDCJG07}.  Therefore, for bubble rafts, one typically
reports an operational definition of the gas area fraction, $\phi$, as
the average cross-sectional area of bubbles that is visible in the
images.  This is typically done by applying a fixed cutoff to the
images to separate pixels inside and outside of the bubbles. One
expects that this method will provide values for $\phi$ that
approximate the definition of the packing fraction used in the bubble
model.  Using this operational definition, we find $\phi = 0.86 \pm
0.04$ for the bubble raft experiments. The error is estimated based on
a range of choices for reasonable cutoff values. As we will show, one
advantage of our studies is that they can provide a method for
calibrating our definition of the gas area fraction for bubble rafts
with the packing fraction for the bubble model.

\begin{figure}
\includegraphics[width=9cm]{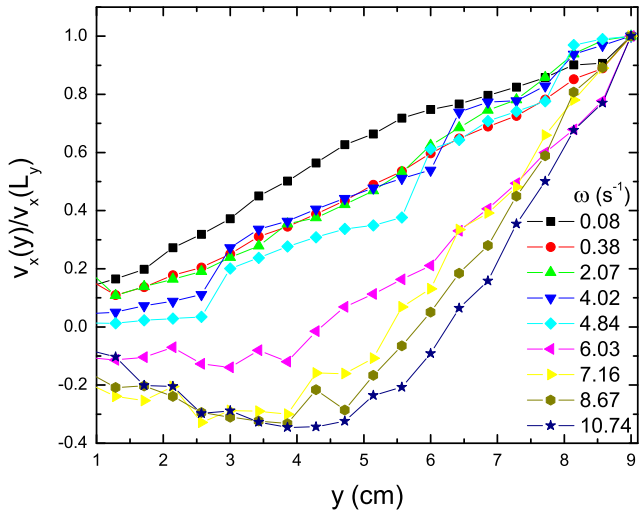}
\caption{(Color online) The normalized average horizontal velocity
$v_x(y)/v_x(L_y)$ (averaged over times when the boundary velocity
is maximum) as a function of distance $y$ from the stationary
bottom wall for several driving frequencies $\omega$ in the bubble
raft experiments. At frequencies $\omega < \omega_{d} \approx
4.0\ {\rm s^{-1}}$, the profiles are roughly linear, while above
this characteristic frequency, liquid-like nonlinear profiles are
observed.} \label{velocity_profile}
\end{figure}

To visualize the bubble motions, a $210 \times 240$ pixel CCD
camera was held above the basin. The floor of the basin was
constructed from glass, which allowed the entire bubble raft to be
illuminated from below by an electroluminescent film (manufactured
by Luminous Film Inc.). A frame rate of $60$ frames per second was
sufficient to capture the bubble motion.  The raw experimental
images were filtered and thresholded to demarcate the spatial
location of each bubble. Further details of the image processing
are provided in Ref.~\cite{wkd06a,wkd07,foamweb}.

To measure the velocity profiles, we divided the system in the
$y$-direction into $\sim 20$ equal-sized rectangular bins between
$y=0$ and $y=L_y$. The instantaneous velocities of bubbles were
then calculated by subtracting the center of mass positions of the
bubbles in consecutive image frames for each bin. We employed a
PIV procedure to measure the time evolution of the bubble
velocities and thus calculate the phase shift relative to the
driving.  The horizontal component of the velocity of the interior
bubbles can be parameterized by the rms velocity $v_{rms} \equiv
\sqrt{\langle v_x^2(y) \rangle}$, frequency of oscillation
$\omega$, and local phase shift $\delta(y)$ with respect to the
moving top boundary.  The horizontal component of the velocity of
the interior bubbles is therefore given by
\begin{equation}
v_x(y,t) = v_{mag}(y) \cos (\omega t - \delta(y)),
\end{equation}
where $v_{mag} = \sqrt{2}v_{rms}$, which allows us to calculate
the local phase shift $\delta(y)$ relative to the moving boundary.
In Sec.~\ref{results} below, we will show results for the total
phase shift $\Delta \equiv \delta(0) - \delta(L_y)$.

In the oscillatory planar shear experimental setup, we are not
able to directly measure the shear stress and thus the flow curve.
However, the flow curve has been measured previously for a bubble
raft with similar parameters (shown in Fig.~\ref{expflowcurve})
\cite{PD03}. We employed the same method as in the simulations to
determine the crossover shear rate ${\dot \gamma}_c \approx 0.065\
{\rm s^{-1}}$ at which the system transitions from the quasistatic
to the power-law flow regime.

\section{Experimental and Simulation Results}
\label{results}

\begin{figure}
\includegraphics[width=9cm]{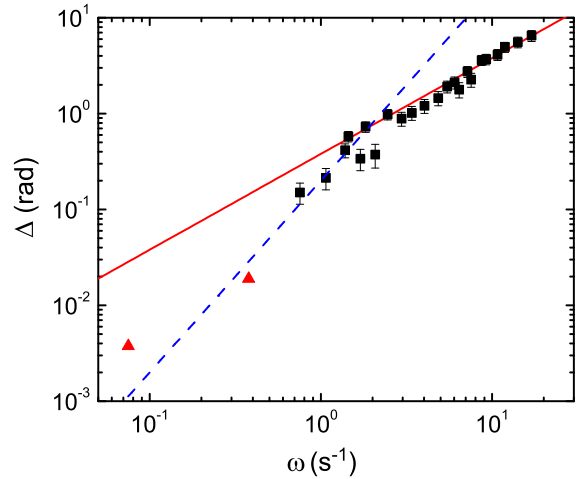}
\caption{(Color online) Total phase difference $\Delta$ plotted as a
function of the driving frequency $\omega$ for the bubble raft
experiments (black squares). The two lowest frequency data points (red
triangles) represent the maximum possible phase shift that can be
measured given the resolution of our experiment. The solid red and
dashed blue lines have slopes $1$ and $2$, respectively.  The
intersection of these two lines provides a rough estimate for the crossover
frequency, $\omega_p \approx 1.9 \pm 0.9\ {\rm s^{-1}}$.}
\label{expphase}
\end{figure}

In Figs.~\ref{velocity_profile} and \ref{expphase}, we show the
results from the experiments in which bubble rafts were subjected
to low amplitude oscillatory planar shear.
Fig.~\ref{velocity_profile} displays the horizontal velocity
profiles $v_x(y)$ (measured at times when the boundary velocity is
maximum as defined in Sec.~\ref{expmeths}), over a range of
driving frequencies.  For this system, we find that the velocity
profiles transition from nearly linear to `liquid-like' nonlinear
near $\omega_{d} \approx 4.0 \pm 0.5\ {\rm s^{-1}}$. Note that at
the lowest driving frequency, it appears that the velocity profile
displays `solid-like' behavior with slight negative curvature
above a linear profile, but this feature is comparable to the size
of the fluctuations. Since this feature is difficult to detect and
the nonlinear liquid-like profiles are more robust in the
experiments, we will focus on measurements of $\omega_d$ instead
of $\omega_{nl}$.

In Fig.~\ref{expphase}, we show measurements of the total phase
difference $\Delta$ between bubbles near the top and bottom boundaries
as a function of the driving frequency. As discussed in
Sec.~\ref{background}, for a viscoelastic material we expect a
crossover in the scaling of the total phase difference as the driving
frequency is increased. For the experiments, we were unable to fully
characterize the crossover behavior due to limits in our ability to
measure the phase shift at low frequencies.  However, using the
maximum possible values of the total phase shift at the lowest
frequencies red triangles in Fig.~\ref{expphase}, we can set a
lower-limit on the exponent for the low-frequency scaling regime.
This gives us a transition from $\Delta(\omega) \sim \omega^n$, with
$n \geq 2$, at low-frequencies to linear scaling at high
frequencies. Thus, for the bubble raft system, we estimate $\omega_p
\approx 1.9 \pm 0.9\ {\rm s^{-1}}$, and thus $\omega_p < \omega_{d}$.

\begin{figure}
\includegraphics[width=9cm]{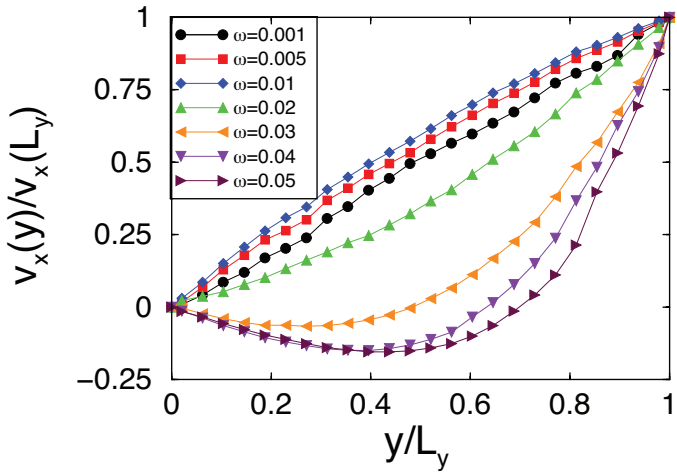}
\caption{(Color online) The normalized average velocity
$v_x(y)/v_x(L_y)$ (averaged over times when the boundary velocity
is maximum) as a function of distance $y$ from the stationary wall
plotted for several driving frequencies $\omega$ in the bubble
model simulations. For $\omega \gtrsim \omega_{d} \approx 0.02$,
the velocity profiles display `liquid-like' nonlinear decay.}
\label{velocity_profile_th}
\end{figure}

\begin{figure}
\includegraphics[width=9cm]{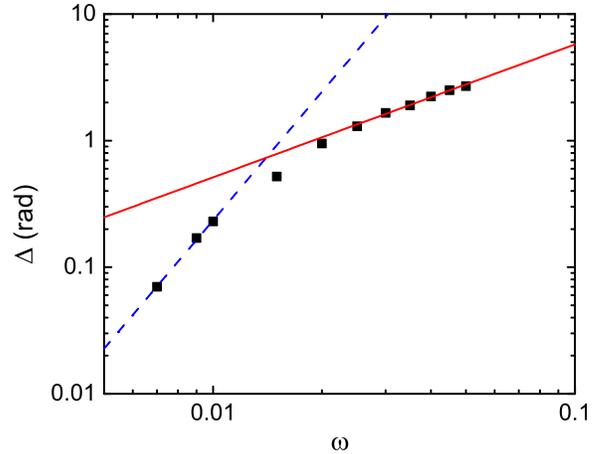}
\caption{(color online) Total phase difference $\Delta$ plotted as
a function of the driving frequency $\omega$ for the bubble model
simulations at $\phi =0.86$ and $b^*=2$. From the intersection of
the solid red and dashed blue lines with slopes $1$ and $3$,
respectively, we estimate $\omega_p \approx 0.015$.}
\label{simphase}
\end{figure}

We also performed simulations of the bubble model undergoing small
amplitude oscillatory planar shear over a range of packing fractions
$\phi$ and damping coefficients $b^*$ to compare to the bubble raft
experiments. The trend for each set of $\phi$ and $b^*$ was similar:
low frequency profiles are linear with a nonzero total phase shift
followed by a transition to nonlinear liquid-like profiles when
$\omega > \omega_{d}$. In addition, the crossover in the scaling
behavior of $\Delta$ versus $\omega$ was observed.  This
characteristic behavior is highlighted in
Figs.~\ref{velocity_profile_th} and ~\ref{simphase}, which show the
velocity profiles and total phase shift for the bubble model with
$\phi = 0.86$ and $b^* = 2$. We find $\omega_d \approx 0.02$ and
$\omega_p \approx 0.015$ for this set of parameters. In the
low-frequency limit, we find $\Delta (\omega) \sim \omega^n$, with $n
\approx 2.8 \pm 0.3$ (blue dashed line in Fig.~\ref{simphase}), which
is similar to the scaling predicted for the Kelvin-Voigt model
\cite{nonlinear}.

Figure~\ref{freqsum} summarizes the simulation data (black squares)
for the transition to liquid-like nonlinear profiles ($\omega_{d}$)
and the crossover frequency ($\omega_p$).  We show both characteristic
frequencies $\omega_d$ and $\omega_p$ at fixed $\phi=0.86$ as a
function of $b^*$ including the under- and overdamped regimes and at
fixed $b^* = 2$ as a function of $\phi$.  To directly compare the
results in experiments and simulations, we normalize $\omega_d$ and
$\omega_p$ by $\omega_c = 2\pi\dot{\gamma}_c$ obtained from the steady
planar shear flow curves at each $b$ and $\phi$. The experimental
results are also displayed in Fig.~\ref{freqsum}; they are represented
as solid lines since $\phi$ and $b^*$ are not known precisely for the
bubble raft experiments.  The simulations indicate that at fixed
$\phi$, the $b^*$ dependence of $\omega_d/\omega_c$ and
$\omega_p/\omega_c$ is fairly weak as the dissipation switches from
underdamped to overdamped.  At fixed $b^*$, both $\omega_d/\omega_c$
and $\omega_p/\omega_c$ increase with packing fraction, if we exclude
the first point at $\phi=0.85$.

\section{Conclusion}
\label{summary}

We find that the bubble model captures all of the qualitative features
of the response of the bubble rafts to low-amplitude oscillatory
planar shear flow.  The key elements of the response include: 1. A
regime with {\it linear} velocity profiles, but a nonzero total phase shift
at low frequencies and 2. At the highest frequencies, a regime with
liquid-like {\it nonlinear} profiles and nonzero total phase shift.  In
addition, in the low frequency regime, there is a crossover in the
scaling of total phase shift as a function of the driving
frequency. The bubble model reproduces each of these distinctive
features found in the bubble rafts.

\begin{figure}
\includegraphics[width=9cm]{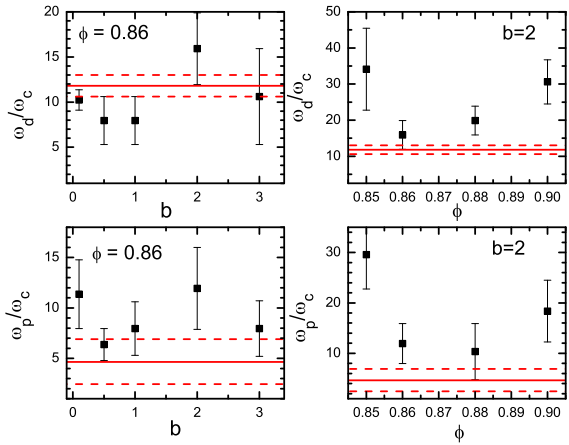}
\caption{(Color online) Summary of the measurements of the normalized
frequencies $\omega_{d}/\omega_c$ (top panels) and
$\omega_{p}/\omega_c$ (bottom panels) for the bubble model simulations
(solid squares) and bubble raft experiments (solid lines).  $\omega_d$
and $\omega_p$ are normalized by $\omega_c$, which is the frequency at
which steady planar shear flows crossover from highly-fluidized to
quasistatic behavior.  For the simulations, we show
$\omega_d/\omega_c$ and $\omega_p/\omega_c$ as a function of $b^*$ at
fixed $\phi = 0.86$ (left panels) and as a function of $\phi$ at fixed
$b^* = 2$ (right panels).  Because it is difficult to define $\phi$
and $b^*$ precisely in experiments, the results for the bubble rafts
are presented as solid lines (in between dashed lines, which give
error bars for the experimental measurements) to identify the range of
$\phi$ and $b^*$ that best fit the experiments.} \label{freqsum}
\end{figure}

The quantitative agreement shown in Fig.~\ref{freqsum} between the
simulations and the experiments is also encouraging. In the range of
parameters that we expect to correspond most closely to the
experiments ($\phi \approx 0.86$ and $b^* > \sqrt{2}$), we find
agreement to within error bars between the experiment and simulations
for the characteristic frequency $\omega_d$ at which the system
transitions from linear to liquid-like nonlinear velocity profiles,
and $\omega_p$, which characterizes the change in scaling of the total
phase shift.

The bubble model is often used to characterize highly-fluidized flows
where T1 bubble rearrangement events occur since it can quantify the
statistics of these rearrangement events.  It has not been employed as
often to quantitatively study slow, dense flows where bubble
rearrangements are rare since the results can depend on the
dissipation model.  One of the unique aspects of this study is that it
allows a direct comparison between the bubble model and bubble raft
experiments in the regime where T1 events and other large-scale bubble
rearrangements {\em do not occur}, which enables stringent tests of
the dissipation mechanism in the bubble model.  The observed
qualitative and quantitative agreement between simulation and
experiment provides evidence that the bubble model provides a
faithful, yet simple description of dissipation between liquid walls
in bubble rafts.

Our focus on the low-amplitude oscillatory shear regime, in the
absence of T1 events, provides important insights into the
response of foams to applied stress. First, we find that the
bubble rafts clearly exhibit dissipative behavior despite the
absence of T1 events. This can only be due to motion in the fluid
films, and emphasizes that further quantitative studies of
film-level dissipation mechanisms are crucial to understanding
foam rheology \cite{dtgal08}. Second, studies of the response of
3D foams and emulsions to oscillatory shear suggest that there are
important differences between the two systems in the low-amplitude
regime \cite{SD99,CHR98,KSA88}. Thus, studying the low-amplitude
regime will highlight key differences in the mechanical response
of a variety of soft glassy materials.  Finally, in this article
we presented novel results that demonstrate the existence of two
additional time (or frequency) scales associated with the response
of foam to applied shear: $\omega_p$ and $\omega_d$ (or
$\omega_{nl}$).  Thus, we expect that these characteristic time
scales will also play a key role in determining the flow curve and
velocity profiles for foams undergoing steady planar shear flow.

A quantitative comparison between the bubble raft experiments and
bubble model simulations also provides a method to calibrate the
gas area-fraction of the bubble rafts, which is notoriously
difficult to measure in foam experiments.  Of the three frequently
used quasi-two dimensional experimental setups (bubble rafts,
bubble rafts with a top glass plate, and bubbles confined between
two solid surfaces), it is most difficult to define the area
fraction in the bubble rafts. By combining the bubble model
simulations and Surface Evolver computations \cite{B96,CJ08}, it
will be possible to define an effective area fraction that will
allow direct comparison between bubble raft experiments and other
quasi two-dimensional glassy and jammed systems confined to
surfaces and thin films.  In the future, we will perform
additional experiments over a range of packing fractions,
surfactants that yield bubbles with varied elastic properties, and
liquid viscosities to investigate whether the predictions of the
simple viscoelastic model in Sec.~\ref{background} for the $G$-
and $\eta-$dependence of $\omega_{nl}$ and $\omega_p$ are valid
for the bubble rafts.

\begin{acknowledgments}
Financial support from the Department of Energy [DE-FG02-03ED46071
(MD), DE-FG02-05ER46199 (NX), and DE-FG02-03ER46088 (NX)], NSF
[DMR-0448838 (CSO) and CBET-0625149 (CSO)], and the Institute for
Complex Adaptive Matter (KK) is gratefully acknowledged.
\end{acknowledgments}


\begin{thebibliography}{40}
\bibitem[{non()}]{nonlinear}
\bibinfo{note}{In the bubble model simulations, we have performed preliminary
  measurements of the frequency-dependent storage and loss moduli, $G'(\omega)$
  and $G''(\omega)$. We find that the linear response regime occurs for $A <
  0.5 \sigma$. Despite this, we find that the scaling behavior of $\omega_p$
  with frequency is consistent with the Kelvin-Voigt model.}
\end{thebibliography}

\end{document}